\documentclass[aps,pre,twocolumn,groupedaddress,amsmath,amssymb,nofootinbib]{revtex4-1}

\usepackage{graphicx,color}
\usepackage{caption}
\usepackage{subcaption}
\usepackage{amsmath}
\usepackage{epstopdf}
\usepackage{soul}
\usepackage{tabularx}
\newcommand{\D}{\mathrm{d}}
\newcommand{\e}{\mathrm{e}}

\newcommand{\be}{\begin{equation}}
\newcommand{\ee}{\end{equation}}
\newcommand{\bea}{\begin{eqnarray}}
\newcommand{\eea}{\end{eqnarray}}
\newcommand{\ba} {\begin{align} }
\newcommand{\ea} {\end{align} }
\newcommand{\eps}{\varepsilon}

\newcommand{\kbt}{k_{\mathrm{B}}T}

\newcommand{\lb}{l_{{\rm B}}}
\newcommand{\ld}{\lambda_{{\rm D}}}
\newcommand{\kd}{\kappa_{{\rm D}}}
\newcommand{\vecr}{\boldsymbol{r}}

\newcommand{\vecq}{\boldsymbol{q}}

\begin{document}
%%%%%%%%%%%%%%%%

%\baselineskip=24pt

%%%%%%%%%%%%%%%%%%%%%%%%%%%%%%%%%%%%%%%%%%%%%%%
\title{Screening length for finite-size ions in concentrated electrolytes }

\author{Ram M. Adar$^{1}$, Samuel A. Safran$^{2}$, Haim Diamant$^3$, David Andelman$^1$}
\email{andelman@post.tau.ac.il}
\affiliation{$^1$Raymond and Beverly Sackler School of Physics and Astronomy\\ Tel Aviv
University, Ramat Aviv, Tel Aviv 69978, Israel\\
$^2$Department of Chemical and Biological Physics, Weizmann Institute of Science, Rehovot 7610001, Israel\\
$^3$Raymond and Beverly Sackler School of Chemistry\\ Tel Aviv
University, Ramat Aviv, Tel Aviv 69978, Israel
}

%%%%%%%%%%%%%%%%%%%%%%%
\begin{abstract}
The classical Debye-H\"uckel (DH) theory clearly accounts for the origin of screening in electrolyte solutions and works rather well for dilute electrolyte solutions. While the Debye screening length decreases with the ion concentration and is independent of ion size, recent surface-force measurements imply that for concentrated solutions, the screening length exhibits an opposite trend; it increases with ion concentration and depends on the ionic size.
The screening length is usually defined by the response of the electrolyte solution to a test charge, but can equivalently be derived from the charge-charge correlation function. By going beyond DH theory, we predict the effects of ion size on the charge-charge correlation function. A simple modification of the Coulomb interaction kernel to account for the excluded volume of neighboring ions yields a non-monotonic dependence of the screening length (correlation length) on the ionic concentration, as well as damped charge oscillations for high concentrations.
\end{abstract}
%%%%%%%%%%%%%%%%%%%%%%

%%%%%%%%%%%%%%%%%%%%%%
\maketitle
%%%%%%%%%%%%%%%%%%%%%%%

%%%%%%%%%%%%% sec: Intro %%%%%%%%%%%%%%
\section{Introduction}
\label{sec1}
%%%%%%%%%%%%%%%%%%%%%%%%%%%%%%%%%%%%%%
Ionic solutions can be found in a wide range of biological and electrochemical systems~\cite{VO,Israelachvily,David95,SafinyaBook,SamBook,Levincor} and are studied for both their fundamental properties and industrial applications. A key feature of ionic solutions, captured already within the seminal work of Debye and H\"uckel (DH)~\cite{DH1,DH2}, is the screening of electrostatic interactions between charged objects immersed in solution~\cite{VO,Israelachvily,David95,SafinyaBook,SamBook,Levincor,DH1,DH2}, characterized by the Debye screening length, $\ld$.

For a solution of monovalent cations and anions of bulk concentration $n_s$ in a solvent of dielectric constant $\eps$, the Debye screening length is given by $\ld\equiv1/\kd=1/\sqrt{8\pi\lb n_s}$, where  $\lb=e^2/\left(4\pi\eps \kbt\right)$ is the Bjerrum length (SI units), $e$ is the elementary charge and $\kbt$ the thermal energy. The Debye length describes not only the screened potential due to external charges, but also the spatial decay of charge-charge correlations that arise from thermal fluctuations in the ionic concentration. Therefore, $\ld$ can also be regarded as the DH charge-charge correlation length.

The Debye length and the DH limiting theory are independent of the ionic diameter. This is reasonable for dilute solutions, but not for more concentrated solutions with a non-negligible volume fraction of ions. In addition, the DH charge-correlation function for finite-sized ions violates the Stillinger-Lovett second-moment condition~\cite{SL1,SL2,Martin88}, originating from the perfect long-range shielding of the electric field within a conducting medium.

Several theories for finite-sized ions~\cite{Kirkwood34,Kirkwood54,Outhwaite69,Outhwaite70,waisman72,Rasiah72,Hall91,Attard 1993, Carvalho94, Caccamo96, Lee96, Lee97, Varela03, Janecek09,Girotto17, Kjellander2018} that improve upon the DH one have been suggested. For example, Lee and Fisher~\cite{Lee97} introduced a generalized DH (GDH) model based on a Debye charging process of charge oscillations. A common feature emerging from these theories is an oscillatory regime of the charge-charge correlation function at moderately high ionic concentrations. Such oscillations were predicted much earlier by Kirkwood~\cite{Kirkwood34}, and the crossover at the onset of oscillations is known as the Kirkwood line. We note that charge oscillations emerge also from several Ginzburg-Landau-type theories that are used, for example, to describe the charge density of ionic liquids in the proximity of charged surfaces \cite{Bazant11,Storey12,Fedorov14,Limmer15,Gavish16,Gavish18}.

 What rekindled the interest in this problem are recent surface-force apparatus (SFA) experiments on several concentrated electrolyte solutions and ionic liquids~\cite{Gebbie15,Smith16,Lee17}. The long-range forces  in these experiments revealed an anomalously large screening length at high ionic concentrations. The measured forces follow a common scaling law~\cite{Smith16,Lee17} for all the electrolytes and ionic liquids used in the experiments. The force decays exponentially with a screening length $1/\kappa$. For low ionic concentrations, $\kd a\ll1$, where $a$ is the ionic diameter, it follows the DH result, $\kappa=\kd$. However, in the limit of $\kd a\gg1$, it scales differently as $\kappa a\sim\left(\kd a\right)^{-2}$. Hence, $\kappa^{-1}$ {\it increases} linearly with the ionic concentration, $\kappa^{-1}\sim n_s$, as opposed to $\kd^{-1}\sim n_s^{-1/2}$.

The above $\kappa^{-1}\sim n_s$ scaling law, inferred from SFA experiments at high ionic concentrations,  was interpreted in Ref.~\cite{Lee17,Lee17b} in terms of solvent molecules that act as defects in a salt crystal, as compared to dilute electrolytes. However, since this elegant picture is based on defects in the crystalline state, which occurs for aqueous NaCl solutions above $6$\,M~\cite{CRC}, while the experiments show an increase in the screening length even around $1$\,M, the understanding of the fluid state is still incomplete. The scaling law has motivated several other recent theoretical works~\cite{Rotenberg18,Kinjal18,Coupette18}, but has not yet been fully understood. While these works rely on different assumptions and yield slightly different results, all of them introduce in a similar way some non-Coulomb short-range interactions between the ions (and possibly, the solvent). These interactions become significant at moderately high concentrations, where the average separation between ions is comparable with the ionic diameter.

In the present work we suggest an alternative approach. The ionic finite size is taken into account by a straightforward modification of the Coulomb interaction. We introduce two such possible modifications: either an interaction that is restricted to separations larger than the ionic diameter, or an interaction between finite-sized ions with a charge distribution, referred to as the internal charge-density~\cite{Frydel13,Frydel16}. A general expression for the charge-charge correlation function is derived analytically, by considering Gaussian charge-density fluctuations in bulk electrolytes.

The outline of our paper is as follows. In Sec.~\ref{sec2}, we present our model and focus in Sec.~\ref{ssec2a} on the modification of the Coulomb interaction for finite-size ions. A general result for the charge-charge correlation function is derived in Sec.~\ref{sec3}, followed by a description of the different regimes of the correlation length in Sec.~\ref{ssec3a}. The dilute and concentrated electrolyte limits are discussed, respectively, in Secs.~\ref{ssec3b} and \ref{ssec3c}. Finally, we relate our findings to experiments and to previous theories, and provide some concluding remarks in Sec.~\ref{sec4}.

%%%%%%%%%%%%%%sec: Model%%%%%%%%%%%%%%%
\section{Model}
\label{sec2}
%%%%%%%%%%%%%%%%%%%%%%%%%%%%%%%%%%%%%%%

%setup
Consider a monovalent electrolyte of bulk concentration $n_s$ at a constant temperature $T$.
%For simplicity, we assume that the cations and anions are both spheres of diameter $a$.
The solvent is modeled as a homogeneous medium with dielectric constant $\eps$. For simplicity, we assume that both ionic species have a diameter $a$ and are spherical-symmetric. Consequently, the interaction between two ions depends only on the ions' separation and not on their orientations.

We focus on properties of the bulk electrolyte, far from any charged objects and surfaces. Furthermore, it is assumed that the electrolyte is far from any liquid-liquid critical point, where the solution phase-separates into two electrolytes of different concentrations~\cite{Lee96}. Here, this assumption means that fluctuations in the ionic concentration can be neglected. This is in accordance with the experimental setup described in Ref.~\cite{Smith16}, where no inhomogeneity was observed.

The internal energy, $U$, has two contributions: a short-range steric
interaction, $U_{\rm sr}$, and a long-range electrostatic one, $U_{\rm el}$.
For symmetric cations and anions, the steric term depends only on the total
ionic concentration, $n=n_++n_-$, whose spatial average is $2n_s$. The electrostatic term, on the other
hand, originates from fluctuations in the number-density {\it difference},
$\rho=n_+-n_-$, whose spatial average is zero, due to electro-neutrality. These fluctuations contribute an
electrostatic energy,
%%%%%%%%%%%%
\begin{align}
\label{eq1}
U_{{\rm el}}&=\frac{\kbt}{2}\int\D^3r\,\D^3r'\,\rho(\vecr)v\left(\vecr-\vecr'\right)\rho(\vecr'),
\end{align}
where $v\left(\vecr\right)$ is the dimensionless electrostatic interaction kernel (units of $\kbt$). Due to the finite size of the ions, $v$ is different than the standard (dimensionless) Coulomb kernel, $v_{\rm C}(\vecr)=\lb/r$, where $\lb$ is the Bjerrum length. This modification of the Coulomb interaction lies at the heart of our work. Possible forms of $v$ are described in detail in Sec.~\ref{ssec2a}.

The free energy, $F=U-TS$, consists of the above energies, as well as the ion mixing entropy, $S$. Up to quadratic order in $\rho$, it is given by
%%%%%%%%%%%%
\begin{align}
\label{eq2}
F\left[n(\vecr),\rho(\vecr)\right]&=F_0[n(\vecr)]+\frac{\kbt}{4
n_s}\int\D^3r\,\rho^2(\vecr)\nonumber\\
&+\frac{\kbt}{2}\int\D^3r\,\D^3r'\,\rho(\vecr)v\left(\vecr-\vecr'\right)\rho(\vecr').
\end{align}
%%%%%%%%%%%%
The first term on the right-hand-side of Eq.~(\ref{eq2}) accounts for the free
energy of a solution of uncharged spheres, while the second term corresponds to
the entropic contribution of small $\rho$ fluctuations, and the third term is
the electrostatic energy of Eq.~(\ref{eq1}).

We note that such an expansion of the free energy is adequate especially for relatively weak electrostatic interactions, as a result of a combination of monovalent ions, high temperatures and high dielectric constants. Such a combination corresponds to Bjerrum lengths that are small and comparable with the ionic diameters. This is indeed the case, for example, for the aqueous NaCl solution used for the surface-force experiments of Ref.~\cite{Smith16}. For such a solution, the Bjerrum length, $\lb\simeq 0.7$\,nm, is approximately the same as the average hydrated ionic diameter~\cite{Nightingale59}.

%%%%%%%%%%%%ssec%%%%%%%%%%%%%%%%%%%
\subsection{Modified electrostatic interaction }
\label{ssec2a}
%%%%%%%%%%%%%%%%%%%%%%%%%%%%%%%%%%%%
It is evident that the quadratic expansion of the free energy in Eq.~(\ref{eq2}) does not couple the fluctuations of the local ionic concentration, $n(\vecr)$, with the charge-density, $\rho(\vecr)$, due to the symmetry between cations and
anions. As a result, the short-range repulsion (that can result from either hard or soft cores) does not enter directly any calculation of electrostatic properties, such as the charge-charge correlation function (see Sec.~\ref{sec3}). However, the contribution of  electrostatics to the free energy is clearly affected by the short-range repulsions. Namely, they preclude microstates where the ion separations are very small and for which the Coulombic interactions are very strong.

The nonphysical contributions of such microstates to the free energy become especially important for concentrated electrolytes with typically small inter-ion separations. In order to prevent or at least to reduce these contributions and better estimate the electrostatic part of the free energy, we consider modified interactions that are weak for small inter-ionic distances, $r<a$, and coincide with the Coulombic one for $r>a$. We refer to the length scale $a$ as the ionic diameter, which can correspond to the bare diameter or hydrated size.

We consider two ways in which the ionic interaction can be modified. First, it is possible to restrict the Coulomb interaction to ionic separations of $r>a$ with a cutoff at $r=a$, given by $v_{\rm co}(r)=\Theta\left(r-a\right)\lb/r$, (for example, see Ref.~\cite{Coupette18}) where $\Theta(x)$ is the Heaviside function. It is useful to rewrite this function in Fourier space. We denote the Fourier transform of a function $f(\vecr)$ as $\widetilde{f}(\vecq)=\int\D^3r\,f(\vecr)\exp\left(-i\vecq\cdot\vecr\right)$, and find that
%%%%%
\be
\label{eq3}
\widetilde{v}_{\rm co}(q)=\frac{4\pi\lb}{q^2}\cos\left(qa\right).
\ee
%%%%%%%
In Fourier space, the minimal distance cutoff in the kernel reduces to a cosine term, while the Coulomb interaction is restored in the $a\to0$ limit.

An alternative approach is to consider ions with an internal charge-density, $e w(r)$. The resulting interaction kernel in Fourier space is
%%%%%%%%
\be
\label{eq4}
\widetilde{v}(q)=\frac{4\pi\lb}{q^2}\widetilde{w}^2\left(qa\right).
\ee
%%%%%%%
The internal charge-density gives rise in Fourier space to the form factor, $\widetilde{w}^2$. As the ions are monovalent, the charge density satisfies $\widetilde{w}(0)=1$, and the Coulomb interaction is restored in the limit $a\to0$.

We focus on the two simplest internal charge densities, $w(r)$, with spherical symmetry: a homogeneous spherical shell, $w_{\rm sh}$, and a homogeneous sphere $w_{\rm sp}$, given by
%%%%%%%%
\be
\label{eq5}
w_{\rm sh} (r)=\frac{1}{\pi a^2}\delta\left(r-\frac{a}{2}\right), ~~w_{\rm sp} (r)=\frac{6}{\pi a^3}\Theta\left(\frac{a}{2}-r\right),
\ee
%%%%%%%
where $\delta(x)$ is the Dirac delta function. The Fourier transform of these charge densities is given in terms of spherical Bessel functions, $j_{n}(x)$, according to
%%%%%%%%
\be
\label{eq6}
\widetilde{w}_{\rm sh} (q)=j_0\left(\frac{qa}{2}\right),~~ \widetilde{w}_{\rm sp} (q)=\frac{6}{qa}j_1\left(\frac{qa}{2}\right).
\ee
%%%%%%%

\begin{figure}[ht]
\centering
\includegraphics[width=0.85\columnwidth]{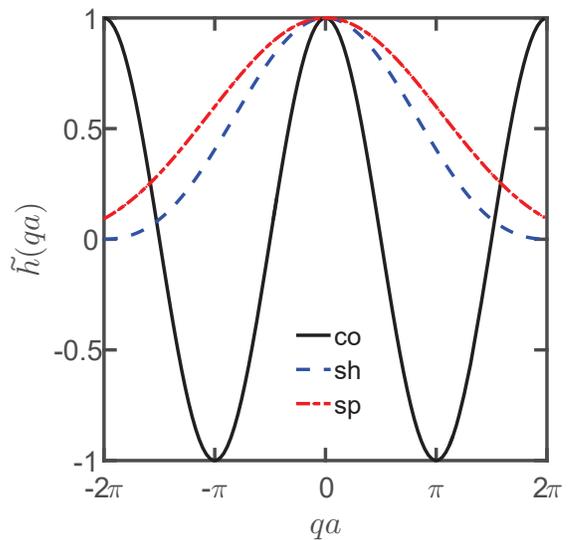}
\caption{(Color online) The $\widetilde{h}$ function as a function of $qa$, for three modified interactions considered in Sec.~\ref{ssec2a} according to Eq.~(\ref{eq8}):  a minimal distance cutoff (co, solid black line), an internal charge-density of a homogeneous spherical shell (sh, dot-dashed red line), and an internal charge-density of a homogeneous sphere (sp, dashed blue line). All functions are even and satisfy $\widetilde{h}(0)=1$.
}
\label{fig1}
\end{figure}

Reviewing Eqs.~(\ref{eq3}) and (\ref{eq4}), we identify a general form of the modified interaction,
%%%%%%%%%
\be
\label{eq7}
\widetilde{v}(q)=\frac{4\pi\lb}{q^2}\widetilde{h}\left(qa\right).
\ee
%%%%%%%%%
We note that $\widetilde{h}$ is an even function, because of the assumed radial symmetry of the interaction, and it satisfies  $\widetilde{h}(0)=1$, reducing to the standard Coulomb interaction for point-like ions. In the three examples above, the different $\widetilde{h}$ functions in Fourier space are given by
%%%%%%%%%
\begin{align}
\label{eq8}
\widetilde{h}_{\rm co}(x)&=\cos(x),\nonumber\\
\widetilde{h}_{\rm sh}(x)&=j_0^2\left(\frac{x}{2}\right),\nonumber\\
\widetilde{h}_{\rm sp}(x)&=\left[\frac{6}{x}j_1\left(\frac{x}{2}\right)\right]^2.
\end{align}
%%%%%%%%%
These three functions are plotted in Fig.~\ref{fig1}.
%%%%%%%%%%%%%%%fig1%%%%%%%%%%%%%
%%%%%%%%%%%%%%%%%%%%%%%%%%%%%%%%%

%%%%%%%%%%sec: Correlation function %%%%%%%%%%%%%
\section{Charge-density correlation function}
\label{sec3}
%%%%%%%%%%%%%%%%%%%%%%%%%%%%%%%%%%%%%%%
The free energy of Eq.~(\ref{eq2}) can be written in Fourier space in terms of the modified interaction of Eq.~(\ref{eq7}), and reads
%%%%%%%%%%
\be
\label{eq9}
F=F_0(n_s)+\frac{\kbt}{4n_s}\int\frac{\D^3q}{\left(2\pi\right)^3}\frac{q^2+\kd^2\widetilde{h}\left(qa\right)}{q^2}\left|\widetilde{\rho}(\vecq)\right|^2,
\ee
%%%%%%%%%%%%
where $1/\kd=1/\sqrt{8\pi\lb n_s}$ is the Debye length. The equipartition over fluctuation modes yields the Fourier transform of the charge-charge correlation function, $\widetilde{G}(q)$,
%%%%%%
\be
\label{eq11}
\widetilde{G}(q)=2n_se^2\frac{q^2}{q^2+\kd^2\widetilde{h}(qa)}.
\ee
%%%%%%

The correlation function of Eq.~(\ref{eq11}) is our main result. It reduces to the DH result in the limit of vanishing ionic diameter ($\widetilde{h}=1$). Furthermore, it coincides for any $a>0$ with the DH result up to quadratic order in $q$ with
%%%%%%
\be
\label{eq12}
\widetilde{G}(qa\ll1)\approx \frac{2n_s e^2}{\kd^2}q^2.
\ee
%%%%%%
This $q$ expansion to leading order satisfies two important conditions. The vanishing zeroth order ($\widetilde{G}(0)=0$) expresses electroneutrality, while the second-order coefficient satisfies the Stillinger-Lovett second-moment condition~\cite{SL1,SL2,Martin88}. We emphasize that $G$ is the two-point correlation function, which originates only from Gaussian fluctuations of the charge density, $\rho$, similarly to the DH correlation function.

The real-space behavior of the correlation function is obtained from the inverse Fourier transform of Eq.~(\ref{eq11}). Its spatial dependence is given by exponential terms,  determined by the poles of $\widetilde{G}(q)$ [Eq.~(\ref{eq11})] that solve the following equation:
%%%%%%%%%%%%%%%
\be
\label{eq12b}
q_0^2+\kd^2\widetilde{h}(q_0a)=0.
\ee
%%%%%%%%%%%%%%%%%%
   These poles are generally complex numbers, and at large distances, the correlation function can be approximated by the contribution of the pole whose imaginary part is closest to the real axis from above. We denote this pole as $q_0=\omega+i\kappa$ with $\kappa>0$, and find that
%%%%%%%%%%%%%%%%
\be
\label{eq13}
G(r\gg a)\approx A \frac{\e^{-\kappa r}}{r}\cos\left(\omega r+\phi\right).
\ee
%%%%%%
where $A$ is the amplitude and $\phi$ the phase. Both $A$ and $\phi$ can be obtained analytically, but we rather focus on $\kappa$ and $\omega$, {\it i.e.}, the value of the pole itself. An analysis of the pole captures the qualitative behaviour of the bulk electrolyte.

%%%%%%%%%%%%ssec%%%%%%%%%%%%%%%%%%%
\subsection{Correlation length }
\label{ssec3a}
\begin{figure}[ht]
\centering
\includegraphics[width=0.85\columnwidth]{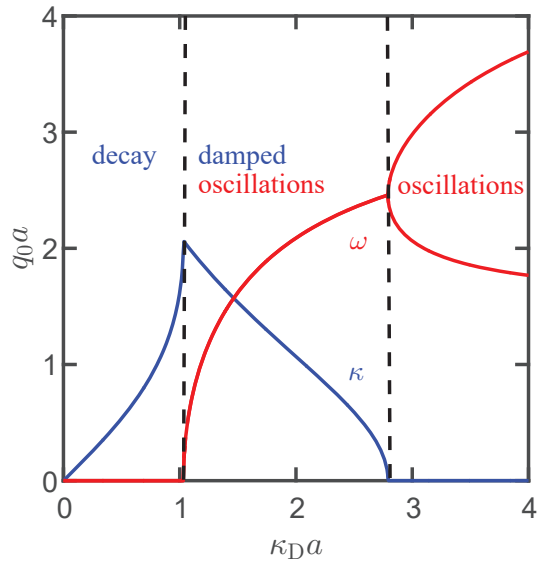}
\caption{(Color online) Inverse decay length $\kappa$ (blue) and oscillation wavenumber $\omega$ (red) for the cutoff modified interaction $\left(\widetilde{h}=\widetilde{h}_{\rm co}\right)$. The left dashed line marks the Kirkwood line, $\kd^\ast$, beyond which damped charge-density oscillations are formed. Beyond the right dashed line, oscillations no longer decay, which signifies the formation of long-range order according to this model. }
\label{fig2}
\end{figure}

\begin{figure}[ht]
\centering
\includegraphics[width=0.85\columnwidth]{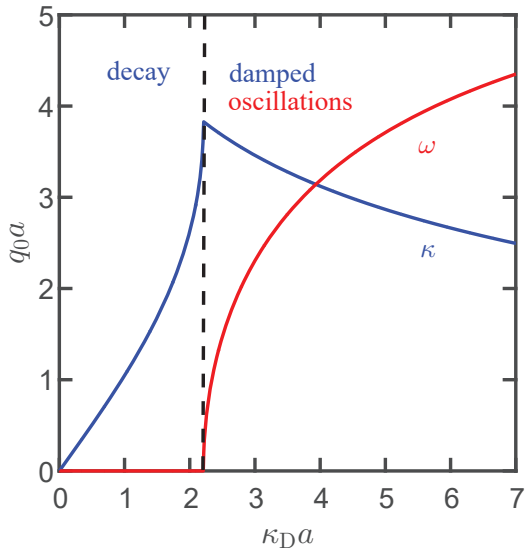}
\caption{(Color online)  Inverse decay length $\kappa$ (blue) and oscillation wavenumber $\omega$ (red) for an internal ionic charge-density on a spherical shell  $\left(\widetilde{h}=\widetilde{h}_{\rm sh}\right)$.  The  dashed line marks the Kirkwood line, $\kd^\ast$, beyond which damped charge-density oscillations are formed. There is no transition to pure oscillations (long-range order) in this case.
}
\label{fig3}
\end{figure}

%%%%%%%%%%%%%%%%%%%%%%%%%%%%%%%%%%%%
As is evident from Eq.~(\ref{eq13}), the real part of the pole, ${\rm Re}(q_0)=\omega$, defines a wavenumber of charge-density oscillations, while the imaginary part, ${\rm Im}(q_0)=\kappa$, defines the charge-density inverse screening length. A purely imaginary $q_0=i\kappa$ describes a gas-like phase with a finite correlation length, $1/\kappa$. A general complex $q_0$ corresponds to a liquid-like disordered phase, with finite-ranged charge-density oscillations. Finally, a purely real $q_0=\omega$ infers a long-range order of alternating positive and negative charges having a wavelength $2\pi/\omega$. Plots for $q_0$ as a function of $\kd$ [Eq.~(\ref{eq12b})] for the different modified Coulomb interactions are presented in Figs. \ref{fig2} and \ref{fig3}.

%%%%%%%%%%%%%%%fig2%%%%%%%%%%%%%
%%%%%%%%%%%%%%%%%%%%%%%%%%%%%%%%%

The inverse decay length and oscillation wavenumber that correspond to the cutoff interaction, $\widetilde{h}_{\rm co}$, are depicted in Fig.~\ref{fig2}. Inserting this interaction in Eq.~(\ref{eq12b}) restores the exact same criterion for the decay length found by Kirkwood~\cite{Kirkwood34,Kirkwood54}. For very dilute solutions $\left(\kd a\ll1\right)$, the DH result is restored with $q_0=i\kd$. For higher concentrations, the correlation is purely decaying, and the inverse decay length, $\kappa$, increases with $\kd$ and slightly differs from the DH result. This behavior persists up to the Kirkwood line, $\kd=\kd^{\ast}$ (left dashed line), where the pole has a non-zero real part, $\omega>0$, and the finite-ranged charge-density fluctuations become damped oscillatory.

Beyond the Kirkwood line, the wavenumber increases as the inverse decay length decreases. For high $\kd$ values, the inverse decay length vanishes for the cutoff modified interaction (right dashed line), and the pole is purely real, corresponding to a long-range order. This behavior, as is plotted in Fig.~\ref{fig2}, qualitatively restores also the results of Lee and Fisher~\cite{Lee97}, obtained using the GDH theory. We note that such a long-range order is different than a solid salt crystal and is expected to be unstable at high concentrations. This unphysical high-concentration limit is resolved by considering different modified interactions.

%%%%%%%%%%%%%%%fig3%%%%%%%%%%%%%
%%%%%%%%%%%%%%%%%%%%%%%%%%%%%%%%%

The inverse decay length and oscillation wavenumber that correspond to the modified interaction of an internal charge-density distributed on a spherical shell, $\widetilde{h}_{\rm sh}$, is depicted in Fig.~\ref{fig3}. Similarly to the cutoff interaction of Fig.~\ref{fig2}, $q_0$ is purely imaginary for low ionic concentrations, and oscillations start to occur at the Kirkwood value, $\kd=\kd^\ast$ (dashed line), whose value is larger than in the cutoff case of Fig.~\ref{fig2}. However, unlike the case plotted in Fig.~\ref{fig2}, here the disordered phase persists for arbitrarily high ionic concentrations, and no long-range order is formed. Rather, the inverse decay length gradually decays with $\kd$. Finally, for the modified interaction due to a homogeneous spherical ionic charge-density ($\widetilde{h}_{\rm sp}$) of each ion, the results are qualitatively similar to those plotted in Fig.~\ref{fig3} and are not presented explicitly.

It is evident that charge-charge correlations in dilute electrolytes exhibit the same dependence on $\kd$, regardless of the exact form of the modified interaction. This is sensible, because the typical separation between ions in a dilute solution exceeds the ionic diameter, making finite-size effects less significant. This universal behavior in the dilute limit is analyzed below in Sec.~\ref{ssec3b}. In the opposite limit of concentrated electrolyte, Figs. \ref{fig2} and \ref{fig3} demonstrate qualitatively different results for the two types of modified interactions. This concentrated limit is further explored below in Sec.~\ref{ssec3c}.
%%%%%%%%%%%%ssec%%%%%%%%%%%%%%%%%%%
\subsection{Dilute electrolyte limit }
\label{ssec3b}
%%%%%%%%%%%%%%%%%%%%%%%%%%%%%%%%%%%%
For low ionic concentrations, similarly to the DH result, the inverse decay length is expected to be small. Therefore, the poles of the correlation function in $q$-space occur at small values. They can be found from an expansion of $\widetilde{h}$ in powers of $qa$, $\widetilde{h}(x)\approx 1+Ax^2+Bx^4$, where $A<0$ and $B>0$. Such second and fourth order terms are in accordance with the modified interactions considered in this work.

Substituting the small $qa$ expansion of $\widetilde{h}$ in Eq.~(\ref{eq11}) yields the following equation for the pole, $q_0$:
%%%%%%%%%%%
\be
\label{eq14}
\left(q_0a\right)^4+\left(q_0a\right)^2\left(\frac{1}{B\left(\kd a\right)^2}+\frac{A}{B}\right)+\frac{1}{B}=0.
\ee
%%%%%%%%%%
 %Inverting this equation yields
%%%%%%%%%%%
%\be
%\label{eq15}
%\left(q_0 a\right)^2=\frac{-\left(\left(\kd a\right)^{-2}+A\right)+\sqrt{\left(\left(\kd a\right)^{-2}+A\right)^2-4B}}{2B}.
%\ee
%%%%%%%%%%
For very low ionic concentrations ($\kd a\ll 1$), Eq.~(\ref{eq14}) simplifies to $q_0^2\kd^{-2}+1=0$, restoring the DH result $q_0=i\kd$. The Kirkwood value $\kd^\ast$ is found as the root of the discriminant of Eq.~(\ref{eq14}), {\it i.e.},
%%%%%%%%
\be
\label{eq16}
\kd^\ast a=\left(\sqrt{4B}-A\right)^{-1/2}.
\ee
%%%%%%%%%
At the Kirkwood value, the inverse decay length obtains its maximal value, $B^{-1/4}$.

We note that a low $q$-expansion of the correlation function and its poles is rather general and can also be performed within other frameworks. Interestingly, the inverse decay length, obtained from the low-concentration $q_0$ of Eq.~(\ref{eq14}), agrees with experimental results taken from Ref.~\cite{Smith16} for low and moderate concentrations. This is explored further in the Appendix~\ref{appA}.

\subsection{Concentrated-electrolyte limit}
\label{ssec3c}

At high ionic concentrations, $\kd a \gg 1$, the oscillation wavenumber, $\omega={\rm Re}(q_0)$, saturates, while the inverse decay length, $\kappa={\rm Im}(q_0)$, gradually decreases (Figs. \ref{fig2} and \ref{fig3}). We assume that $\kappa$ decays algebraically with the ionic concentration, according to $\kappa a=b \left(\kd a\right)^{-\alpha}$, where $b $ and $\alpha$ are positive numbers. In order to determine the value of the decay exponent, $\alpha$, this expansion is substituted in the left-hand side of Eq.~(\ref{eq12b}), and the pole of the correlation function is found by equating separately the real and imaginary parts of all orders of $\kd a$ to zero.

\begin{figure}[ht]
\centering
\includegraphics[width=0.85\columnwidth]{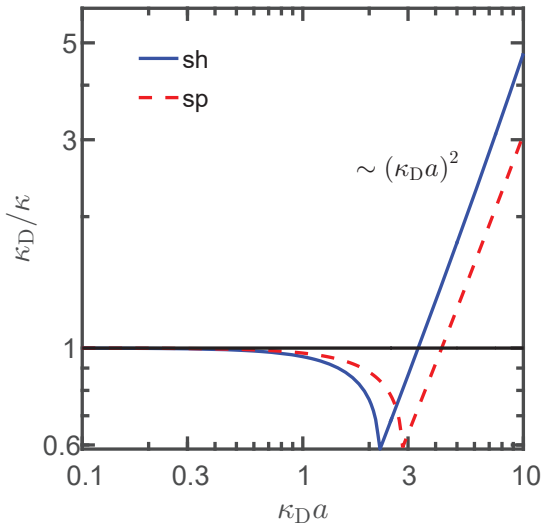}
\caption{(Color online) The decay length normalized by the Debye length, $\kd/\kappa$, as function of $\kd a $ (log-log scale) for two possible internal charge densities: a homogeneously charged spherical shell (solid blue line) or homogeneously charged sphere (dashed red line). For low concentrations, $\kappa=\kd$, as determined by DH theory. In the concentrated electrolyte limit, the inverse decay length scales according to $\kd/\kappa\sim\left(\kd a\right)^2$.
}
\label{fig4}
\end{figure}

The oscillation wavenumber is determined by solving for the highest order term in $\kd$, $\kd^2\widetilde{h}\left(\omega a\right)=0$. The next-order term vanishes for
%%%%%%%%%%
\be
\kd^{2-\alpha}\widetilde{h}'\left(\omega a\right)b =0.
\label{eqa1}
\ee
%%%%%%%%%%%%%%%%%%
 We distinguish between two possible solutions. For $b =0$, the inverse decay length is identically zero, as is the case for the cutoff interaction with $\widetilde{h}_{\rm co}$. Alternatively, $\omega a$ can be a degenerate root of $\widetilde{h}$, such that $\widetilde{h}'\left(\omega a\right)=0$ and  $\kappa$ decays gradually. This is the case for the internal charge densities, where $\widetilde{h}\sim \widetilde{w }^2\left(qa\right)$. We focus on this latter case, and find the appropriate $\alpha$ from the next order term,
%%%%%%%%%%%
\be
\label{eqa2}
\left(\omega a\right)^2-\frac{1}{2}\widetilde{h}''\left(\omega a\right)b ^2\left(\kd a\right)^{2-2\alpha}=0,
\ee
%%%%%%%%%%%%
%As $k^2$ is positive, we require that $n=2\left(2m-1\right)$, where $m$ is a natural number. Consequently, $\alpha$ is given by $\alpha=1/\left(2m-1\right)$. The maximal value of $\alpha$, therefore, is $\alpha=1$. The next largest possible value is $\alpha=1/3$. The amplitude, $\delta$ is given by $\delta=\left[n!k^2/h^{(n)}(k)\right]^{1/n}$. In particular, for the internal charge densities considered here we find that $\alpha=1$.
resulting in $\alpha=1$. This result can be written in the form
%%%%%%%%%%
\be
\label{eqa3}
\left.\frac{\kd}{\kappa}\right|_{\kd a\gg1}\sim\left(\kd a\right)^2.
\ee
%%%%%%%%%%
The above relation is in accordance with our numerical calculations of $\kappa$ for $\widetilde{h}_{\rm sh}$ and $\widetilde{h}_{\rm sp}$, as is demonstrated in Fig.~\ref{fig4}.  The above scaling does not coincide with the master curve produced in Ref.~\cite{Smith16} from experimentally measured decay lengths, where another scaling law is proposed: $\kd/\kappa\sim\left(\kd a\right)^3$.

%%%%%%%%%%%%%%%fig4%%%%%%%%%%%%%
%%%%%%%%%%%%%%%%%%%%%%%%%%%%%%%%%
%%%%%%%sec: Discussion %%%%%%%%%%%%%%%%%%%%%%%%%%%%%%%%%%%%%%%%%%%%%%
\section{Discussion }
\label{sec4}
%%%%%%%%%%%%%%%%%%%%%%%%%%%%%%%%%%%%%%%

We propose a simple theory to account for the finite ion size and its effect on charge-charge correlations. Rather than adding higher order terms that couple the charge density to the ionic concentration by short-range repulsions, some of the short-range part of the Coulombic interaction is subtracted. This modification of the two-body Coulomb interaction is simply expressed in Fourier space by the function $\widetilde{h}\left(qa\right)$. Such a general modification can be easily extended and can also be interpreted in terms of non-local electrostatic effects~\cite{Bazant11,Storey12,Storey12}.

  It is important to note that the short-range modification of the Coulombic kernel affects the charge-density correlation only due to microstates where some ions are separated by distances smaller than $a$. This is the reason that the modified electrostatic energy due to an internal ionic charge-density [Eq.~(\ref{eq4})] was used in previous works to describe soft penetrable ions, such as polyelectrolytes~\cite{HansenA,HansenB,HansenC}. It was shown that the internal charge-density can result in charge inversion and over-screening~\cite{Frydel13}, as well as a Kirkwood crossover~\cite{Frydel16}.

  In the case of simple ions, penetrability is possible if $a$ is interpreted, for example, as the hydrated ion diameter, rather than the bare hardcore that is impenetrable due Pauli's exclusion principle. In addition, we find this description useful for concentrated electrolytes, because the smearing of the charge diminishes the nonphysical magnitude of the interactions at small ionic separations, $r<a$. Such a procedure is also used to bypass divergent self-energies in field-theoretical descriptions~\cite{Wang10}. Similarly, it was shown that the mean spherical approximation (MSA) can be interpreted in terms of the interaction between such spherical charged shells~\cite{Roth16}.

%We described how the modified interaction results in damped charge-density oscillations for high ionic concentrations. In the case of the small distance cutoff, oscillations become pure for even higher concentrations and a long-range order is formed. The same qualitative result was obtained using the GDH theory~\cite{Lee97} %as well as more recent theories\cite{Rothenberg19} , indicating that it is a robust result. Unlike the GDH theory, our framework can be extended in a straight-forward manner. In particular, considering interactions that originate from an internal charge density, we show that no pure oscillations are formed.
 The modified interaction between ions with an internal charge density has a notable physical feature. For high ionic concentrations, rather than pure charge oscillations, it yields damped oscillations (Fig.~\ref{fig3}), which decay according to $\kappa a\sim1/\kd a$.  This scaling was also observed in MC simulations for closely-packed ionic liquids~\cite{Kinjal18}, by varying the temperature for a fixed ionic concentration. However, the scaling was observed only close to the Kirkwood line, and was not extended to much larger $\kd$ values. In this large $\kd$ limit, the charge-frustrated Ising model of Ref.~\cite{Kinjal18} predicts pure charge oscillations.
% A MSA calculation, on the other hand, yields $\kappa \sim1/\sqrt{\kd}$.

Recent SFA experiments~\cite{Smith16} suggest that the electrostatic forces decay with an inverse decay length that decreases more rapidly with the ionic concentration. The relation  $\kappa a \sim1/\left(\kd a\right)^2$ was shown to be satisfied by several concentrated electrolytes and ionic liquids~\cite{Smith16}, which fall on a master curve when $\kappa a$ is plotted as a function of $\kd a$, for all the chemical systems studied. This experimental scaling was also inferred recently~\cite{Gaddam19} from the surface excess of fluorescein in thin films of concentrated ionic solutions. The excess was determined from the detection of fluorescent emission, and it was related to the electrostatic decay length.

One reason for the different scaling in our work and the experimental results lies in the different meanings of  $\kappa$ and $\kd$. The inverse decay length, $\kappa$, in our work is derived from the charge-density correlation length in the {\it bulk} electrolyte, rather than {\it surface} properties or inter-surface forces. Although these length scales are expected to coincide in the limit of large surface separations, surface effects may prevent a clear identification of the measured screening length with the predicted bulk correlation length for charge fluctuations. For example, optical-tweezers experiments on charged colloids in nonpolar solvents~\cite{Bartlett18} suggest that surface charge-regulation may be the origin of the nonmonotonic behavior of the electrostatic decay length observed in that system. We note that future scattering experiments may provide a more direct measure of the bulk correlation length without complications due to surface effects in SFA.

 The inverse Debye length $\kd=\sqrt{2e^2 n_s/ \eps\kbt}$ also deserves attention, due to its dependence on the dielectric constant, $\eps$. In our work, the dielectric constant is independent of the ionic concentration, and the pure solvent value is used. In Ref.~\cite{Smith16}, on the other hand, $\kd$ is given in terms of the dielectric constant of the electrolyte solution, $\eps(n_s)$. The static dielectric constant of electrolytes decreases with concentration for simple salts in water~\cite{BarthelBook,Hasted48,Adar18}, due to excluded solvent volume and electrostatic correlations between the solvent and solute~\cite{Adar18}. For concentrated electrolytes, we have found that $\kappa\sim 1/\kd\sim\sqrt{\eps}$. The decrement of the dielectric constant, $\eps(n_s)$, thus suggests a further decrement of $\kappa$. However, this effect is not strong enough in order to solely explain the experimental scaling of $\kappa\sim 1/\kd^2$.

Another plausible explanation for the difference between our predictions and the measured scaling are electrostatic correlations that lie outside the scope of our Gaussian framework. Extreme correlations in the form of an ionic crystal were considered in Refs.~\cite{Lee17,Lee17b}. We believe that in the fluid state (which is relevant for aqueous solutions of NaCl $<6$\,M~\cite{CRC}, where anomalous screening is already observed), a more physical picture involves ionic clusters (or blobs) of  partially correlated ions. Our framework can be used in the future to describe the interaction between such clusters.

We note that our theory cannot be extended to arbitrarily high ionic concentrations within the above concentrated electrolyte limit. At sufficiently high concentrations, the electrolyte approaches a critical point, where it phase-separates into two electrolytes of different concentrations~\cite{Lee96}. Close to this phase transition, large concentration $(n)$ fluctuations occur~\cite{Lee96}, and the present formulation for a homogeneous electrolyte must be refined. This is evident, for example, from the term  $\sim\rho^2/n_s$ in Eq.~(\ref{eq2}), which can be strongly affected by fluctuations where the ion density is far from its average ({\it e.g.}, small local values of  $n_s$).  However, we note that such fluctuations were not observed in the above mentioned  experiments.

For high concentrations, ion pairing can also become significant~\cite{Bjerrum,Zwanikken09,Adar17,Huang18}. Ion pairing was even suggested as a mechanism for the observed under-screening~\cite{Huang18}. However, for simple salts such as NaCl, pairing is not expected to be substantial for concentration of a few molars, as is indicated, for example, by dielectric data~\cite{Adar18}.

Finally, we mention possible future extensions of our theory. One can consider
other modified interactions and, more specifically, other internal-charge form factors. %For example, it is possible to describe charge densities of solvated ions rather than bare ions or to determine the form factor from a variation of the free energy. Performing such a variation for fixed total charge has resulted in qualitatively similar results.
For example, rather than describing simple ions, the form factors can correspond to correlated ion clusters, as was suggested above. Another possibility is to model the solvent explicitly as charge dipoles in order to account for the
dielectric decrement in the presence of ions. An explicit treatment of
solvent is also important for the structural force between surfaces at small
separations, originating from non-electrostatic forces of concentrated
solvent~\cite{Coupette18}.  The compact analytical form of the correlation
function [Eq.~(\ref{eq11})] should be helpful for developing such extensions in
future studies.

 \vskip 0.5cm
{\it Acknowledgments.~}
We thank Martin Bazant, Kinjal Dasbiswas, Andreas H\"artel, Alpha Lee, Susan Perkin, and Phil Pincus for fruitful discussions and suggestions. SAS is grateful for the support of the Israel Science Foundation and the US-Israel Binational Science Foundation. DA acknowledges partial support from the ISF-NSFC (Israel-China) joint program under Grant No. 885/15.
%%%%%%%%%%%%%%%%%%%%%%%%%%%%%%%%%%%%%%%%%%%%%%%%%
%%%%%%%%%%%%%%%%%%%%%%%%%%%%%%%%%%%%%%%%%%%%%%%%%
%Appendix A
\appendix*
\section{The dilute-limit expansion and its comparison to experimental data}
\label{appA}
\begin{figure}[ht]
\centering
\includegraphics[width=0.85\columnwidth]{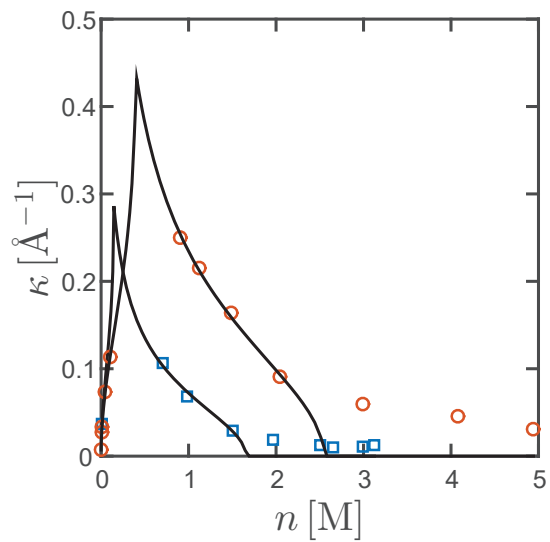}
\caption{(Color online) Inverse decay length inferred from two surface-force experiments~\cite{Smith16}: (i) aqueous NaCl solution (red circles) and (ii) an ionic liquid [C$_{4}$C$_{1}$Pyrr][NTf$_2$]  mixed with propylene carbonate (blue squares). The black curves were fit to the data by solving Eq.~(\ref{eq14}) and treating $A$ and $B$ as fit parameters. For higher concentrations, the decay length of Eq.~(\ref{eq14}) vanishes and undamped charge oscillations occur.
}
\label{fig5}
\end{figure}

%%%%%%%%%%%%%%%%%%%%%%%%%%%%%%%%%%%%%%%%%%%%%%%%%%%%
The inverse decay length, obtained from the low-concentration $q_0$ of Eq.~(\ref{eq14}), agrees with the experimental results of Ref.~\cite{Smith16} for low and moderate concentrations.  Figure~\ref{fig5} presents data from two experiments in Ref.~\cite{Smith16} for the electrostatic decay length of an aqueous NaCl solution (red circles) and an ionic liquid in propylene carbonate (blue squares). The black curves correspond to the solution of Eq.~(\ref{eq14}), with the coefficients $A$ and $B$ treated as fit parameters. For the NaCl solution, the fitted values are $A=-1.94,\,B=0.50$, while for the ionic liquid, they are $A=-1.34,\,B=0.32$.

Negative $A$ values are expected from the Taylor expansion of the $\widetilde{h}$ functions of Eq.~(\ref{eq8}), reflecting how the Coulombic interaction is diminished. Smaller $A$ values correspond to larger values of the Kirkwood line, $\kd^\ast a$. The $B$ values, on the other hand, correspond to the peak of the inverse decay length, which satisfies $\kappa a=B^{-1/4}$. In addition, $a$ values were inferred from x-ray scattering experiments~\cite{Santos11} for the ionic liquid ($a=0.465$ nm) and correspond to the average bare ionic diameter of NaCl ($a=0.27$ nm~\cite{Nightingale59}). At high ionic concentrations, the inverse decay length of Eq.~(\ref{eq14}) vanishes.

We note that a low $q-$expansion of a modified interaction kernel can also be performed within other frameworks. For example, within the Bazant-Storey-Kornyshev~\cite{Bazant11,Storey12} model, the non-local solution permittivity is expanded in powers of the wavenumber, resulting in a modified interaction. The fit to experimental data in Fig.~\ref{fig5}, therefore, does not validate to our theory, but rather demonstrates how a low-$q$ modification of the interaction kernel is in accordance with the experimental data up to moderate ionic concentrations.

%%%%%%%%%%%%%%%fig5%%%%%%%%%%%%%
%%%%%%%%%%%%%%%%%%%%%%%%%%%%%%%%%

\newpage
%%%%%%%%%%%%%%%%%%%%%%%%%%

\end{document}